\documentclass{article}
\usepackage{fortschritte}
                            
\def\be{\begin{equation}}                      
\def\ee{\end{equation}}                       
\def\bea{\begin{eqnarray}}                    
\def\eea{\end{eqnarray}}                       
              
\def\NP{{\it Nucl. Phys.} }                   
\def\PL{{\it Phys. Lett.} }                   
\def\PR{{\it Phys. Rev.} }                    

\newcommand{\ft}[2]{{\textstyle\frac{#1}{#2}}}
\def\der{\partial}

\newcommand{\rmd}  {{\mathrm{d}}}
\newcommand{\Vol}  {\mathrm{Vol}}
\newcommand{\refeq}[1]{(\ref{#1})}
\begin {document} 
\begin{center}
\hfill hep-th/0312195\\
\hfill FSU-TPI-14/03
\end{center}         
\def\titleline{
%
%
%
Space-Time Singularities and the K\"ahler Cone
}
%
%
%
\def\authors{
%
%
%
%
%
L. J\"arv, C. Mayer, T. Mohaupt\footnote{corresponding author E-mail: {\tt mohaupt@tpi.uni-jena.de}}, F. Saueressig                           
}
\def\addresses{
%
%
%
%
Theoretisch-Physikalisches Institut,
Friedrich-Schiller-Universit\"at Jena \\
Max-Wien-Platz 1, D-07743 Jena, Germany \\

}
\def\abstracttext{
%
%
We review recent results on the interplay between 
the five-dimensional space-time and the internal manifold
in Calabi-Yau compactifications
of M-theory. Black string, black hole and domain wall solutions as well 
as Kasner type  cosmologies cannot develop a naked singularity as long as 
the moduli take values inside the K\"ahler cone.
%
}
\large
\makefront

String and M-theory provide mechanisms and effects
which improve on the problem of space-time singularities. 
Besides the resolution of singularities 
through $\alpha'$-effects, the most interesting mechanism is 
what one might call `the intervention of additional modes.'
M-theory is a theory of extended objects, and winding states
of strings and p-branes become massless for special values of 
the moduli parameterising the geometry of the  
compactified dimensions. When describing M-theory in terms of
an effective action, it is crucial
that {\it all} the light modes are taken into account. If one only includes
those states which are massless away from the special points
in the moduli space, then solutions with naked singularities
are as generic as solutions without. However, in the context
of the so-called enhan\c{c}on geometry it was observed that
naked singularities may be artifacts, which disappear
when the dynamics of all relevant M-theory modes is taken into
account \cite{Enhancon}. 
It was soon realised that the same mechanism is at work
in compactifications of M-theory on Calabi-Yau threefolds \cite{KMS,TM1}. 
In this note we review recent progress in establishing 
the absence of naked singularities for certain classes of five-dimensional
space-times in a model-independent way, {\it i.e.}, valid for
compactifications on arbitrary Calabi-Yau threefolds \cite{JMS,MM}. The 
result of this work is that, due to a beautiful interplay between the
internal manifold and space-time, electric and magnetic BPS solutions,
BPS domain wall solutions, and Kasner cosmological 
solutions cannot develop naked singularities as long as the
scalar fields take values in the interior of the M-theory moduli space,
which in the case at hand is the so-called extended
K\"ahler cone of the Calabi-Yau threefold. Moreover, 
the asymptotic behaviour of space-time geometries at the 
boundary of moduli space is determined by the type of degeneration
of the internal manifold. 
Space-time singularities can
occur when the boundary of the extended K\"ahler cone is reached, but
in these cases the description in terms of a five-dimensional effective
action breaks down, because infinitely many M-theory degrees of freedom
become massless. Hence, these are not singularities of specific solutions, but 
rather indications that one needs a completely different description of the system.


The five-dimensional supergravity actions \cite{GST}
considered here contain, besides the supergravity multiplet, 
$n_V$ vector multiplets
and $n_H$ hypermultiplets. The action is fixed once the vector and hypermultiplet scalar manifolds have 
been specified and a gauging has been chosen. The vector multiplet
manifold is a very special real manifold, {\it i.e.}, a 
hypersurface characterised by a homogeneous cubic polynomial,
called the prepotential,
\be
{\cal V}(X) := \ft16\, C_{IJK} X^I X^J X^K \stackrel{!}{=} 1 \;,
\label{PrePot}
\ee
where $I = 0, 1, \ldots, n_V$. The hypermultiplet manifold is
quaternion-K\"ahler.
We will consider the following cases: (i)  actions without gaugings, 
where all  multiplets are gauge-neutral and the scalar potential vanishes
identically, and (ii) actions with a specific gauging, which correspond to 
Calabi-Yau compactifications with internal $G$-flux \cite{M-DW}.


When a five-dimensional supergravity action is 
obtained by the compactification of eleven-dimensional supergravity
on a Calabi-Yau threefold $X$, then it is 
fixed by the geometrical and topological data of the internal
space \cite{11to5}. 
We expand the K\"ahler form $J$ of $X$ in a basis of the second cohomology group
\begin{equation}
  \label{eq:1,1}
  J = \mathcal{Y}^I\omega_I\;,\qquad \langle\omega_I\rangle = H^{1,1}(X)\;,\quad
  n_V = \dim H^{1,1}(X) - 1\;,
\end{equation}
with real moduli $\mathcal{Y}^I$. 
These moduli are related to the scalars $X^I$ of eq.\ (1) by
$\mathcal{Y}^I$~$=$~$V^{1/3}X^I$, where $V$ is the volume of $X$.
We choose a dual basis of four-forms, $\nu^I$, two-cycles, $C_I$, and
four-cycles, $D_I$, by Poincar\'e duality and intersection duality 
respectively,
{\it i.e.,~} $\int_X$~$\nu^I\wedge\omega_J$~$=$~$\delta^I_J$, and
$\int_{C^I}\omega_J$~$=$~$\delta_J^I$.  The coefficients $C_{IJK}$ of
eq.\ (\ref{PrePot}) 
now have the interpretation of triple-intersection numbers,
$C_{IJK}$~$=$~$D_I\circ D_J\circ D_K$, which implies that they are
\emph{integer valued} in contrast to generic five-dimensional 
supergravity actions where \emph{real valued}
constants are allowed.  

The K\"ahler cone of $X$ is defined by the requirement that all
holomorphic curves $C$~$\subset$~$X$, surfaces $S$~$\subset$~$X$, and $X$ itself have
positive volume, {\it i.e.,~}$\int_{C}$~$J$~$>$~$0$, 
$\int_{S}$~$J\wedge J$~$>$~$0$, and $V$~$=$~$\mathcal{V}(\mathcal{Y})$~$>$~$0$.
We consider \emph{primitive}
boundaries, which are codimension-one boundaries where a single 2-cycle,
$C^\star$, collapses. We call a basis {\it adapted}, if the K\"ahler cone
takes the form $\mathcal{Y}^I = \int_{C^{I}} J = \mbox{Vol}(C^I) > 0$.
It is known that away from the so-called cubic cone the K\"ahler cone is 
locally polyhedral, 
whereas at the cubic cone the Calabi-Yau volume $V$ vanishes. 
Therefore we can always choose such an adapted parameterisation, at least
locally. In this basis primitive contractions correspond to blowing down
one of the basic 2-cycles, 
$\mathcal{Y}^{\star} \rightarrow 0$, where $\star \in \{0, \ldots, n_V \}$. 

At primitive boundaries of the K\"ahler cone, the following
contractions can take place \cite{WW}: 
\begin{itemize}
\item type I (``$2\rightarrow0$''): A finite number $n$ of \emph{isolated} 
curves in the homology class  $C^\star$ is blown down
to a set of points, $\Vol(C^\star)=\mathcal{Y}^\star\rightarrow0$. This
results in $n$ charged hypermultiplets becoming massless.
\item type II (``$4\rightarrow0$''): A divisor $D=v^ID_I$ collapses to a set of points:
  $\Vol(D)$~$\propto$~$(\mathcal{Y}^\star)^2$. This results in 
tensionless strings, implying that infinitely many particle-like
states become massless.
\item type III (``$4\rightarrow2$''): A (complex) one-dimensional family of curves sweeps
  out a divisor $D=v^ID_I$. Contracting this family of curves induces 
the collapse
of $D$ into a curve: $\Vol(D)$~$\propto$~$\mathcal{Y}^\star$.
This results in $SU(2)$ gauge symmetry enhancement, possibly 
accompanied by a finite number of massless hypermultiplets in 
the adjoined representation.
\item Cubic cone (``$6\rightarrow4$'', ``$6\rightarrow2$'', ``$6\rightarrow0$''): These
  contractions correspond to $V$~$\propto$~$\mathcal{Y}^\star$,
  $V$~$\propto$~$(\mathcal{Y}^\star)^2$ and $V$~$\propto$~$(\mathcal{Y}^\star)^3$.
No M-theory interpretation is known, and it is clear that the 
description in terms of a five-dimensional effective action breaks down.
\end{itemize}
Boundaries of type I and type III can be crossed into the
K\"ahler cone of a new Calabi-Yau
threefold, which is birationally (and, for type III, even biholomorphically)
equivalent to the original one.  The extended K\"ahler
cone is obtained by enlarging the K\"ahler moduli space at all boundaries of type I.
Enlarging in addition the K\"ahler moduli space at all boundaries of type III
results in
the extended movable cone. However, this second extension 
only adds ``gauge copies'' to the
parameter space (see \cite{MohZag,TM1} for an explanation).
While type I and type III boundaries are ``internal boundaries'' 
of the M-theory
moduli space, type II contractions and the cubic cone 
lead to proper boundaries. 

The metric on the Calabi-Yau K\"ahler moduli space is given by
\cite{Str+Bod}
\begin{equation}
  \label{eq:KC_metric}
  G_{IJ} := \frac{1}{2V}\;\int_X\omega_I\wedge\star\omega_J 
  = -\frac{1}{2}\;
  \frac{\partial}{\partial \mathcal{Y}^I}\frac{\partial}{\partial
  \mathcal{Y}^J}\log \mathcal{V}(\mathcal{Y}) \;.
\end{equation}
This metric is non-degenerate inside the K\"ahler cone. The link between 
curvature singularities of space-time 
and properties of the K\"ahler-cone metric
is provided
by the matrix
\begin{equation}
  M_{IJ} = \frac{1}{2}\;\int_X J\wedge\omega_I\wedge\omega_J =
  \frac{1}{2}\;C_{IJK} \mathcal{Y}^K \;,
\end{equation}
which is related to the K\"ahler-cone metric by \cite{MM}
\begin{equation}
  \label{eq:detGM}
  \det (G_{IJ})  = -\frac{1}{2}\left(\frac{-1}{V}\right)^{\dim H^{1,1}(X)}\det ( M_{IJ} )\;.
\end{equation}
\begin{table}
\begin{tabular*}{\textwidth}{@{\extracolsep{\fill}} lclc}  
    \hline \hline
    \multicolumn{2}{c}{type of boundary} & physics & behaviour of $\det(G_{IJ})$  \\
    \hline
    \hspace*{3mm} internal \hspace*{3mm} & \hspace*{8mm} type I \hspace*{8mm}   &flop transition       &  regular\\
             & type III & gauge symmetry enhancement & regular\\
    \hline
 \hspace*{3mm}    external  & type II  &tensionless strings& zero\\
             & cubic cone &unknown& divergent\\
    \hline \hline
  \end{tabular*}
  \caption{\label{BoundRes} Behaviour of the K\"ahler moduli-space metric at the boundaries of the
    extended K\"ahler  cone. 
  }
\end{table}
Since $G_{IJ}$ is non-degenerate inside the K\"ahler cone, so is $M_{IJ}$.

It remains to analyse the behaviour of $G_{IJ}$ (and $M_{IJ}$) at the primitive 
boundaries of the K\"ahler cone.
As can be seen from eq.\ \refeq{eq:detGM} 
$\det(G_{IJ})$ diverges at the cubic cone ($V=0$).
The three remaining types of boundaries (type I--III) are analysed as follows.
Using eq.\ \refeq{eq:detGM}, for finite and non-zero Calabi-Yau volume $V$,
we are able to infer regularity properties of the K\"ahler-cone metric $G_{IJ}$ from
the matrix $M_{IJ}$ and vice versa: there is a one-to-one map of zero eigenvalues of
$G_{IJ}$ to zero eigenvalues of $M_{IJ}$. If
 $\det(M_{IJ})\big|_{\mathcal{Y}^\star\rightarrow0}\propto(\mathcal{Y}^\star)^n$,
then there are $n$ linearly independent eigenvectors of $M_{IJ}$ (and of $G_{IJ}$)
satisfying
\begin{equation}\label{eq:zeroEVs}
  v^I_{(i)}M_{IJ}\big|_{\mathcal{Y}^\star\rightarrow0} = 0\;,\quad i=1\dots n\;.
\end{equation}
Eq.\ \refeq{eq:zeroEVs} is supposed to hold throughout the face
$\mathcal{Y}^\star$~$=$~$0$.  
In particular, the null
eigenvectors are determined by the triple intersection numbers, only.
This implies that the components of the
eigenvectors can be chosen to be \emph{integer}. 
%
%
Hence, each zero eigenvector $v_{(i)}^I$ defines a divisor
\begin{equation}
  D_{(i)} := v^I_{(i)}D_I\;.
\end{equation}
If there is a holomorphic surface within the homology
class $D_{(i)}$, then its volume vanishes always as $(\mathcal{Y}^\star)^2$ \cite{MM}.
As a consequence, the divisors $D_{(i)}$,
which are associated to null eigenvectors $v_{(i)}$, must perform
a type-II contraction, where $\Vol(D)$~$\propto$~$(\mathcal{Y}^\star)^2$,
rather than a type-III contraction, which is characterised by
$\Vol(D)$~$\propto$~$\mathcal{Y}^*$.
Thus we learn that the moduli-space metric is always regular at 
boundaries of type I and type III, while 
it develops a zero eigenvalue at boundaries of type II.  
Table \ref{BoundRes} summarises our result, which is valid for \emph{all} Calabi-Yau
three-folds \cite{MM}. 
%



We now consider ungauged five-dimensional supergravity actions,
which come from Calabi-Yau compactifications without flux or 
light brane winding states. These theories have string-like magnetic BPS solutions \cite{BlackString}
\be
\rmd s^2 = e^{-U(r)} \Big\{-\rmd t^2 + \rmd z^2 \Big\} 
+ e^{2 U(r)} \Big\{ \rmd r^2 + r^2 \rmd\Omega_{(2)}^2 \Big\}  \;.
\ee
Here, $e^{3 U(r)}$ $=$ ${\cal V}\big(Y(r)\big)$ is determined by 
the rescaled scalar fields $Y^I(r)$~$:=$~$e^{U(r)}X^I(r)=
e^{U(r)} V^{-1/3} \mathcal{Y}^I(r)$, which 
must be harmonic functions with respect to the transverse coordinates.
For single-centered solutions this means
$Y^I(r) = H^I(r) = c^I + \ft{p^I}{r}$, 
where $r$ is the transverse radial coordinate, $c^I$ are determined
by the values of the moduli at transverse infinity, and
$p^I$ are the magnetic charges. The magnetic components of the
gauge fields are proportional to $\der_r Y^I(r)$. 
The Ricci scalar of this metric takes the form
\be
R = - \, r^{-1} e^{-2U} \, \left( 
\ft32 \, r \, (U')^2   + 2 \, r \, U'' +  4 \,  U' \right)\;.
\ee
Since other
curvature invariants take a similar form, it is straightforward
to show that for $r\not=0$ singularities can only occur if either
$e^U \rightarrow 0$ 
or if $U'$ or $U''$ diverge. The latter can happen only if 
derivatives of the scalar fields $Y^I$ diverge \cite{TM1}. For $r\rightarrow 0$
one either approaches a supersymmetric fixed point leading to
a regular event horizon, so that the solution describes a supersymmetric
black string \cite{BlackString},
or the solution becomes singular at $r=0$,
which, after a suitable rescaling, can be treated 
as a limit of the discussion  for $r \not=0$. Using a parameterisation 
adapted to the K\"ahler cone,
we immediately see that the solutions cannot become singular 
inside the K\"ahler cone, where $Y^I(r) > 0$ and all derivatives
are bounded. The singularity $e^{U} \rightarrow 0$ corresponds
to a particular limit of the cubic cone \cite{TM1}: the overall
volume $V$ sits in a hypermultiplet, and therefore it is constant
for magnetic BPS solutions. Eq.\ (\ref{PrePot}) shows 
that some of the $\mathcal{Y}^I$ go to zero, while others go to
infinity, in such a way that the overall volume is kept constant.

%
%

Ungauged five-dimensional supergravity also has electric
BPS solutions \cite{Sabra,BlackString}
\be
\rmd s^2 = - e^{-4 U(r)} \rmd t^2 + e^{2 U(r)} 
\Big\{ \rmd r^2 + r^2 \rmd\Omega_{(3)}^2 \Big\} \;.
\ee
Again, $e^{3U(r)}$~$=$~${\cal V}(Y(r))$ can be expressed in terms 
of rescaled scalar fields 
$Y^I(r)$~$=$~$e^{U(r)}X^I(r)$
$=$
$e^{U(r)}V^{-1/3} \mathcal{Y}^I(r)$, which this
time have to satisfy the generalized stabilization equation
\be
C_{IJK} Y^J(r) Y^K(r) = 2 H_I (r) \;.
\label{GenStabEq}
\ee
Here $H_I(r) = c_I + \ft{q_I}{r^2}$ are harmonic functions of the
transverse coordinate, $c_I$ are determined by the moduli at
infinity, and $q_I$ are the electric charges.
The electric components of the gauge fields are proportional to
$\der_r ( e^{3 U(r)} Y^I(r) )$ and the Ricci scalar takes the form
\be
R = - r^{-1} e^{-2U} 
 \Big( 6 \, r \, (U')^2 + 2\, r \, U'' + 6 \, U' \Big) \;.
\ee
As the other curvature invariants again have the same structure, we see that
curvature singularities can only occur if either $e^U \rightarrow 0$, which 
corresponds
to a particular limit of the cubic cone, 
or if $U'$ or $U''$ diverges. The latter
only happens if derivatives of the scalar fields $Y^I$ diverge.  In general, one
cannot solve eq.\ (\ref{GenStabEq}) for $Y^I(r)$ in terms of the
harmonic functions. But fortunately, one does not need the explicit solution, because
the occurrence of space-time singularities is controlled by the
metric of the K\"ahler cone. To see this one uses that the quantities $Y^I$ and
$\mathcal{Y}^I$ are proportional.
The factor of proportionality is an algebraic function of the $H_I$ and
$Y^I$, which is finite and non-vanishing 
for $r>0$. By differentiating the generalized
stabilisation equation (\ref{GenStabEq}) once, we obtain the relation 
$Y^{\prime I}$~$=$~$\frac{1}{2}\tilde{M}^{IJ}H'_J$, where 
$\tilde{M}_{IJ} = \ft12 C_{IJK} Y^K$ is a rescaled version of $M_{IJ}$.
We can then use 
eq.\ \refeq{eq:detGM} to make the 
connection to the metric of the K\"ahler cone. Similar arguments
apply to the second derivatives of the moduli. 
As a consequence solutions are regular
inside the K\"ahler cone and on type I and type III boundaries, 
whereas they become singular on type II boundaries and on the cubic cone. 
In a naive supergravity treatment one finds generic electric BPS
solutions with naked singularities, 
where `generic' means that one does not need
to tune parameters to get a singular solution \cite{KMS,TM1}. 
The results of \cite{MM}
guarantee that such singularities are artifacts, if they are not 
related to the cubic cone
or to a type-II contraction. `Faked singularities' result from not 
taking into account the threshold corrections of the states which
become massless on internal boundaries  (type I or type III)
of the M-theory moduli space. This proves that the 
enhan\c{c}on-like mechanism, which was observed to be present 
in particular models in \cite{KMS,TM1} works for arbitrary Calabi-Yau
threefolds. 

For technical reasons, we assumed $r>0$, so that the harmonic functions (and
all their  derivatives) are finite. 
If the limit $r \rightarrow 0$ is regular, the solution 
approaches a supersymmetric fixed point, which is the event
horizon of an extremal black hole \cite{Sabra,BlackString}. If not, 
then the boundary  of the extended K\"ahler cone is reached in this limit.
%
%
%
%

We now turn to BPS domain walls, which are solutions of 
gauged five-dimensional supergravity. We only consider the
particular gauging which describes the bulk dynamics of 
five-dimensional heterotic M-theory, which can be obtained by dimensional reduction 
on a Calabi-Yau threefold with internal $G$-flux. The 
domain walls take the form \cite{M-DW}
\be
\rmd s^2 = e^{2U(y)} \Big\{ -\rmd t^2 + (\rmd x^1)^2 + (\rmd x^2)^2 + (\rmd x^3)^2 \Big\}
+ e^{8 U(y)} \rmd y^2 \;.
\label{DW}
\ee 
This time there are no gauge fields along the five-dimensional space-time, but
the gauging induces a non-trivial scalar potential, which acts as a source of
stress energy. As in the previous cases, $e^{3U(y)}$~$=$~${\cal
  V}(Y(y))$ is determined in terms of rescaled scalar fields
$Y^I(y)$~$:=$ $e^{U(y)}X^I(y)$ $=$
$e^{U(y)} V^{-1/3}(y) \mathcal{Y}^I(y)$ $=$
$e^{-U(y)}\mathcal{Y}^I(y)$.
Note that this time the hypermultiplet scalar $V$, which parameterises
the overall volume of the Calabi-Yau space, is not constant, but given
by $V(y)$ $=$ ${\cal V}(Y(y))^2$ $=$ $e^{6U(y)}$.  
As in the electric case, the
fields $Y^I(y) $ have to satisfy the generalized stabilization equation
\be
C_{IJK} Y^J (y) Y^K(y) = 2 H_I(y)\;,
\ee
where now $H_I(y)$~$=$~$b_I+a_I y$ are harmonic functions with
respect to the single transverse coordinate. The Ricci scalar is
\be
R = 4 \; e^{-8U}\Big( 3 \big(U'\big)^2 - 2 U''\Big)\;.
\ee
Despite minor differences in details, the result of the analysis of 
curvature singularities is the same as for electric BPS 
solutions \cite{MM}.
%
%
%
%

Let us finally discuss cosmological solutions of ungauged
five-dimensional supergravity \cite{JMS}. In contrast to the previous
cases, these solutions are time-dependent and not BPS.
For simplicity, we only consider five-dimensional flat
FRW cosmologies, though the result also applies to more general
Kasner solutions. It turns out to be convenient
to keep a non-trivial lapse function:
%
\be
\rmd s^2 = - e^{8 U(t)} \rmd t^2 + e^{2 U(t)} \Big\{ (\rmd x^1)^2 + (\rmd x^2)^2 + (\rmd x^3)^2
+ (\rmd y)^2 \Big\} \, .
\label{FRW}
\ee
Note that metrics of this type are related to domain walls
(\ref{DW}) by a double Wick rotation. The Ricci scalar of this metric is 
given by
\be
R = - 4\; e^{-8 U} \Big( 3 \dot{U}^2 - 2 \ddot{U} \Big) \;.
\ee
With this particular choice of lapse
function, the scalar equation of motion becomes the standard 
geodesic equation on the scalar manifold,
\be
\ddot{\Phi}^{\Sigma} + \Gamma^{\Sigma}_{\Lambda \Theta}
\dot{\Phi}^{\Lambda} \dot{\Phi}^{\Theta} = 0 \;,
\ee
with time $t$ as affine parameter. Here $\Phi^{\Sigma}$ denotes
all scalar fields, both from vector and from hypermultiplets.
Since there is no scalar potential, the only source of stress-energy
is the kinetic energy $T = \ft12 g_{\Sigma \Lambda} \dot{\Phi}^{\Sigma}
\dot{\Phi}^{\Lambda}$, where $g_{\Sigma \Lambda}$ is the direct
sum of the metrics of the vector and hypermultiplet scalar manifolds.
In this case the link between the space-time geometry and the moduli is
provided by Friedmann's equation, $4 \dot{U}^2 = T $, and $\ddot{U} = 0$, 
where the last equation is a direct consequence of the conservation of $T$ along geodesics. The latter also implies that finite-energy solutions
cannot reach a boundary associated with the cubic cone in 
finite time, as the
derivative of the scalar field associated with the diverging eigenvalue of the
metric has to vanish asymptotically.  Note that this statement 
does not only apply 
to the time coordinate $t$ in (\ref{FRW}), but also to the cosmological 
time $\tau$, 
$\rmd \tau = e^{4U(t)} \rmd t$.
Moreover, solutions are manifestly 
non-singular, as long as the metric on the M-theory moduli space is regular.
The behaviour of solutions on all types of boundaries is discussed in detail
in \cite{JMS}. There we 
also consider cosmological solutions of a particular type of gauged
five-dimensional supergravity \cite{FlopPap}, which
is obtained by ``integrating in'' the charged hypermultiplets which become
massless in a flop transition. 
%
%

Our results obtained for black string, black hole and domain 
wall solutions as well as Kasner cosmologies support the conjecture that there is a general mechanism, which avoids space-time singularities
in M-theory compactifications through the interplay with
the internal dimensions. A natural next step is to consider
type II compactifications on Calabi-Yau threefolds, which
amounts to adding $\alpha'$-corrections to the setup
considered in this paper. More ambitiously, one can
try to obtain analogous results without specifying the
space-time geometry explicitly. This requires to link
space-time and moduli space through
estimates and inequalities, rather than equalities.
One immediate question is whether there is a link between
energy inequalities  in space-time and properties of the
moduli space metric.
%
%
%
%
%


\begin{thebibliography}{77}
%
\bibitem{Enhancon}
C.V.\ Johnson, A.W.\ Peet and J.\ Polchinski, \PR D {\bf 61}
(2000) 086001, [arXiv:hep-th/9911161].
C.V.\ Johnson, R.C.\ Myers, A.W.\ Peet and S.F.\ Ross, 
\PR D {\bf 64} (2001) 106001, [arXiv:hep-th/0105077].

\bibitem{KMS}
R.\ Kallosh, T.\ Mohaupt and M.\ Shmakova, {\it J. Math. Phys.}
{\bf 42} (2001) 3071, [arXiv:hep-th/0010271].

\bibitem{TM1}
T.\ Mohaupt, {\it Fortsch. Phys.} {\bf 51} (2003) 787,
[arXiv:hep-th/0212200].

\bibitem{JMS}
L.\ J\"arv, T.\ Mohaupt and F.\ Saueressig, 
[arXiv:hep-th/0310174], [arXiv:hep-th/0311016].

\bibitem{MM}
C.\ Mayer and T.\ Mohaupt, [arXiv:hep-th/0312008].
  
\bibitem{GST}
M.\ G\"unaydin, G.\ Sierra and P.K.\ Townsend, 
\NP B {\bf 242} (1984) 244.
B.\ de Wit and A.\ Van Proeyen, {\it Commun. 
Math. Phys.} {\bf 149} (1992) 307, [arXiv:hep-th/9112027].
A.\ Cersole and G.\ Dall'Agata, \NP B {\bf 585} (2000) 143,
[arXiv:hep-th/0004111].

\bibitem{M-DW}
A.\ Lukas, B.A.\ Ovrut, K.S.\ Stelle and D.\ Waldram,
\PR D {\bf 59} (1999) 086001, [arXiv:hep-th/9803235],
\NP B {\bf 552} (1999) 246, [arXiv:hep-th/9806051].

\bibitem{FlopPap}
L.\ J\"arv, T.\ Mohaupt and F.\ Saueressig, [arXiv:hep-th/0310173].

\bibitem{11to5}
A.C.\ Cadavid, A.\ Ceresole, R.\ D'Auria and S.\ Ferrara, 
\PL B {\bf 357} (1995) 76, [arXiv:hep-th/9506144].
G.\ Papadopoulos and P.K.\ Townsend, \PL B {\bf 357}
(1995) 300, [arXiv:hep-th/9506150].

\bibitem{WW}
P.M.H.\ Wilson, {\it Invent. Math.} {\bf 107} (1992) 561,
{\it erratum ibid.} {\bf 114} (1993) 231. E.\ Witten, \NP {\bf 471}
(1996) 195, [arXiv:hep-th/9603150].

\bibitem{MohZag}
T.\ Mohaupt and M.\ Zagermann, JHEP 12 (2001) 026,
[arXiv:hep-th/0109055].


\bibitem{Str+Bod}
A.\ Strominger, {\it Phys. Rev. Lett.} {\bf 55}
(1985) 2547. 
M.\ Bodner, A.C.\ Cadavid and S.\ Ferrara,
{\it Class. Quant. Grav.} {\bf 8} (1991) 789.


\bibitem{BlackString}
A.H.\ Chamseddine and W.A.\ Sabra, \PL B {\bf 460} (1999)
63, [arXiv:hep-th/9903046].

\bibitem{Sabra}
W.A.\ Sabra, {\it Mod. Phys. Lett.} A {\bf 13} (1998) 239,
[arXiv:hep-th/9708103].



\end{thebibliography}
\end{document}